\documentclass[preprint,proceedings]{rmaa}

\suppressfulladdresses 

\usepackage{natbib}                 
\usepackage{hyperref}
\usepackage{url}

\usepackage{paralist}

\usepackage{psfrag,color}

\SetYear{2007}
\SetConfTitle{The Nuclear Region, Host Galaxy and Environment of Active Galaxies}

\title{Multiwavelength modeling of TeV AGN observed by H.E.S.S.}

\author{
  J.-P. Lenain\altaffilmark{1}
}

\altaffiltext{1}{LUTH, Observatoire de Paris, CNRS, Universit{\'e} Paris Diderot; 5 Place Jules Janssen, 92190 Meudon, France (\href{mailto:jean-philippe.lenain@obspm.fr}{jean-philippe.lenain@obspm.fr}).}

\shortauthor{Lenain}
\shorttitle{Multiwavelength modeling of TeV AGN observed by H.E.S.S.}

\listofauthors{J.-P. Lenain}
\indexauthor{Lenain, J.-P.}

\abstract{The High Energy Stereoscopic System (H.E.S.S.) experiment, a ground-based $\gamma$-ray \v{C}erenkov telescope array located in Namibia, has now detected many extragalactic objects, which redshifts range from z=0.00183 up to z=0.2, possibly more. With the increasing performances of \v{C}erenkov telescopes, it now becomes possible to probe these objects at small timescales in $\gamma$-ray, allowing the study of regions thought to be very close to the central supermassive black holes. Furthermore, H.E.S.S. has confirmed a $\gamma$-ray emission from M\,87, which is thus the first extragalactic source seen at the TeV range that is not a blazar.
  
  Among blazars, TeV BL\,Lacs are the most challenging objects to test the jet emission models and to shed light on particle acceleration mechanisms. The study of blazars with H.E.S.S. also revealed various temporal behaviors among them. Some objects presents a highly variable X-ray flux with small variation of the $\gamma$-ray, while others show the inverse behavior. The interpretation of such differences is puzzling.
  
  Observations at very high energies also bring indirect measurements of the infrared extragalactic background light (EBL). The interpretation of $\gamma$-ray emission of radiogalaxies such as M\,87 in terms of misaligned blazars and the understanding of the properties of the EBL represent new challenges brought by H.E.S.S. observations of extragalactic sources.
}

\resumen{El experimento High Energy Stereoscopic System (H.E.S.S.), un array de telescopios \v{C}erenkov de rayos $\gamma$ en suelo situado en Namibia, ya ha detectado muchos objetos extragal{\'a}cticos, cuyos corrimientos al rojo var{\'i}an de z=0.00183 a z=0.2, probablemente m{\'a}s. Con el rendimiento creciente de los telescopios \v{C}erenkov, ahora es posible de investigar las peque{\~n}as escalas de tiempo en rayos $\gamma$ de estos objetos, permitiendo el estudio de zonas que pueden situarse muy cercas de los agujeros negros supermasivos centrales. Adem{\'a}s, H.E.S.S. ha confirmado una emisi{\'o}n de rayos $\gamma$ de M\,87, que se convierte entonces en la primera fuente extragal{\'a}ctica vista en TeV energ{\'i}a que no es de blazar.
  
  Entre los blazars, los TeV BL\,Lacs son los objetos m{\'a}s estimulantes para evaluar los modelos de emisi{\'o}n del jet y examinar los procesos de aceleraci{\'o}n de part{\'i}culas. La investigaci{\'o}n de los blazars con H.E.S.S. revelaba tambi{\'e}n distintos comportamientos temporales entre ellos. Unos objetos manifiestan un flujo muy variable en rayos X con peque{\~n}as varaciones de los rayos $\gamma$, mientras que otros muestran el comportamiento inverso. La interpretaci{\'o}n de esas diferencias es desconcertante.
  
  Las observaciones a muy altas energ{\'i}as traen tambi{\'e}n medidas indirectas del fondo infrarrojo extragal{\'a}ctico. La comprensi{\'o}n de la emisi{\'o}n de rayos $\gamma$ de radiogalaxias como M\,87 en cuanto a blazars desalineados, y de las propiedades del fondo infrarrojo extragal{\'a}ctico representan unos de los nuevos desaf{\'i}os propuestos por las observaciones de H.E.S.S. de las fuentes extragal{\'a}cticas.
  
}

\addkeyword{AGN objects: all}
\addkeyword{Galaxies: active}
\addkeyword{Galaxies: individual (M\,87, PKS\,2155$-$304)}

\begin{document}
\maketitle

\section{Detecting Very High Energy $\gamma$-ray from the ground with H.E.S.S.}

\begin{figure} 
  \resizebox{\hsize}{!}{\includegraphics{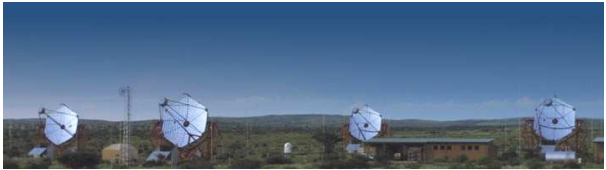}}
  \caption{The H.E.S.S. array located in Namibia.}
  \label{fig:HessT}
\end{figure}

The High Energy Stereoscopic System (H.E.S.S.) experiment\footnote{\url{http://www.mpi-hd.mpg.de/hfm/HESS/}} consists of four imaging atmospheric \v{C}erenkov telescopes (see Figure~\ref{fig:HessT}), fully operating since 2004, located in the Khomas highlands in Namibia. It operates in the very high energy (VHE) domain, in the GeV--TeV range. The analysis of the \v{C}erenkov light of a single air shower allows the determination of the energy of the incident particle. Furthermore, the stereoscopy technique allows a precise determination of the arrival direction, as well as a good discrimination between hadrons and photons\citep[][]{2004APh....22..109A,2006A&A...457..899A}.

\section{TeV blazars}

Blazars are active galactic nuclei (AGNs) with the jet closely aligned with the line of sight, thus having their intensity strongly amplified by relativistic effects. Their spectral energy distribution usually shows two bumps, one in the UV--X range thought to be due to synchrotron emission, and the other one in the X--$\gamma$ range ascribed to inverse Compton effects in leptonic models.

Historically, \citet{1998MNRAS.299..433F} emphasized the existence of two classes of objects: low-peaked BL\,Lac (LBL) and high-peaked BL\,Lac (HBL) for which the synchrotron bump peaks at low ($\sim$10$^{13}$\,Hz), respectively high ($\sim$10$^{16}$\,Hz), frequency. In this framework, the first TeV sources were searched among the HBL \citep{2002A&A...384...56C}, hence all but one AGNs detected by H.E.S.S. belong to this class. M\,87 (FR\,I) and BL\,Lac (LBL) itself are the only non-HBL objects detected in the VHE range up to now by H.E.S.S. and MAGIC respectively \citep[][respectively]{2006Sci...314.1424A,2007ApJ...666L..17A}.

\begin{figure} 
  \resizebox{\hsize}{!}{\includegraphics{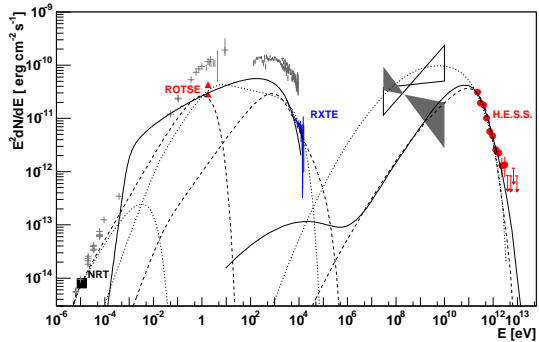}}
  \caption{Spectral energy distribution of PKS\,2155$-$304. Only simultaneous observations are labeled. The solid line is a hadronic model, the dotted and dashed lines are the same leptonic models with different assumptions ({\it from \citet{2005A&A...442..895A}}).}
  \label{fig:PKS2155_MWL}
\end{figure}

Among these sources, different X/$\gamma$ spectral variability patterns can be found. For instance, the X-ray peak observed in H2356$-$309 strongly constrains leptonic models. H2356$-$309 presents low flux variations in X-ray, with rather constant spectral shape, with high flux variations in $\gamma$-ray. On the contrary, PKS\,2005$-$489 has a quasi constant $\gamma$-ray emission along with huge flux variations in X-ray coming with highly varying spectral shape.

\begin{figure} 
  \resizebox{\hsize}{!}{\includegraphics{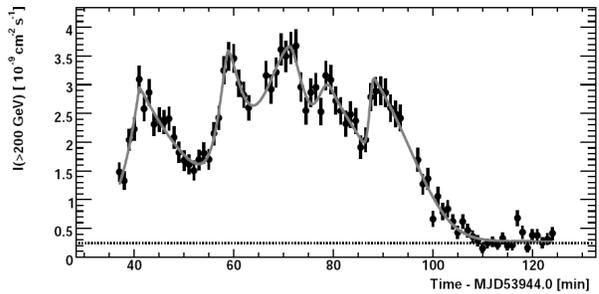}}
  \caption{Lightcurve of PKS\,2155$-$304 on July 28, 2006. Variability is seen down to 2 minutes scale ({\it from \citet{2007ApJ...664L..71A}}).}
  \label{fig:PKS2155_flare}
\end{figure}

PKS\,2155$-$304, a well known variable object, was observed in a long quasi-quiescent state \citep{2005A&A...442..895A} during multiwavelength campaigns in 2003, 2004 and 2005 (see Figure~\ref{fig:PKS2155_MWL}). Observations in July 2006 revealed a very active state over two weeks, with in particular a big flare in VHE up to 15 times the Crab flux with variability timescale of the order of 2 minutes \citep{2007ApJ...664L..71A}. This is the fastest variability ever seen in PKS\,2155$-$304 at any wavelength and in any blazar (see Figure~\ref{fig:PKS2155_flare}).

\section{M\,87 as a misaligned blazar ?}

Between 2003 and 2006, H.E.S.S. observed the FR\,I radiogalaxy M\,87 \citep{2006Sci...314.1424A}, thus confirming the detection of the HEGRA collaboration. Flux variability was observed, without significant spectral variability, with timescales down to 2 days, thus excluding the radio lobes or the extended kiloparsec jet as VHE emitters. It is still unclear whether the emission comes from the nucleus or the knot {\it HST}-1.

\begin{figure} 
  \resizebox{\hsize}{!}{\includegraphics[angle=-90]{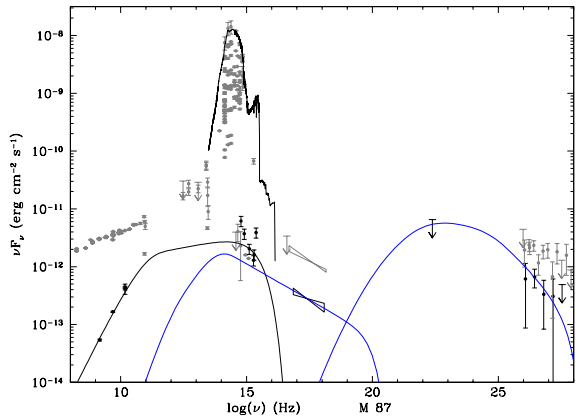}}
  \caption{Spectral energy distribution of M\,87. The black line in the radio shows a model of extended inner jet, while the two lines from IR to VHE show a SSC model to account for the synchrotron and inverse Compton bumps ({\it from \citet{tmp_lenain}}).}
  \label{fig:M87_SED}
\end{figure}

We assume that the VHE emitting zone is located in the broadened jet formation region, predicted by magnetohydrodynamics models and observed in VLBI in the case of M\,87 \citep[see e.g.][]{1999Natur.401..891J}. It is then possible to explain the $\gamma$-ray radiation in terms of synchrotron self-Compton (SSC) processes, as for blazars and with similar electron energy distribution, with rather moderate Lorentz factor, confirming the thought that M\,87 could be a misaligned BL\,Lac object.

The radiation model presented here relies on the previous work by \citet{2001A&A...367..809K}. Here, we assumed that a cap of seven blobs of plasma is located at $\sim$50 gravitational radii from the supermassive black hole, beyond the Alfv{\'e}n surface predicted by magnetohydrodynamic (MHD) models \citep[see e.g.][]{2006MNRAS.368.1561M}. The population of non-thermal electrons of the blobs, while crossing the Alfv{\'e}n surface, is accelerated and then radiates through synchrotron emission  and inverse Compton scattering  of electrons on the synchrotron photons. While radiating in the broadened zone of the jet, some blobs are moving close to the line of sight, with differential Doppler boosting from one blob to the other, allowing to derive parameters similar to those of TeV blazars. Figure~\ref{fig:M87_SED} shows the spectral energy distribution deduced from such a model of M\,87 accounting for the recent VHE H.E.S.S. observations \citep[see][for more details]{tmp_lenain}.

\section{Cosmological implications: The EBL and the $\gamma$ horizon}

The infrared extragalactic background light (EBL) consists of the sum of the starlight and dust emission of the galaxies through the history of the Universe, including starburst and active galaxies. It could also have an important contribution from the first stars, the so-called population~III, which may have formed from primordial gas.

\begin{figure} 
  \resizebox{\hsize}{!}{\includegraphics{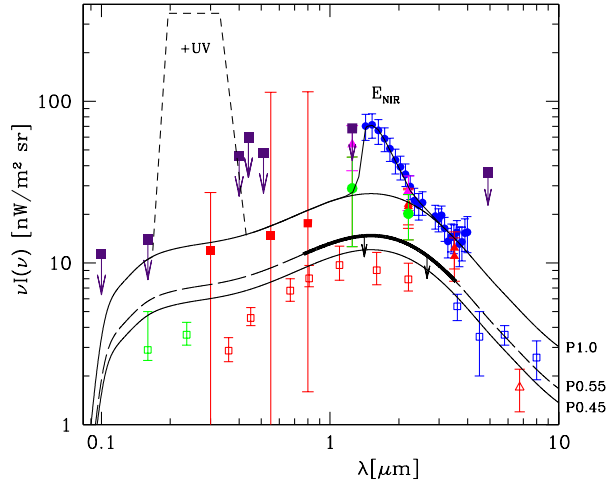}}
  \caption{Spectral energy distribution of the EBL, with the H.E.S.S. upper limit reported ({\it bold black line}) ({\it from \citet{2006Natur.440.1018A}}).}
  \label{fig:ebl_sed}
\end{figure}

While traveling to the earth, VHE photons interact with infrared photons from the extragalactic background light (EBL). The EBL thus absorbs a part of the observed VHE flux. Relying on observations of nearby blazars, it was found that their intrinsic photon index is $\Gamma > 1.5$. It is then possible to indirectly derive measurements of this background through VHE observations of remote ($z \sim 0.2$--$0.3$) blazars \citep{2006Natur.440.1018A} (see Figure~\ref{fig:ebl_sed}).

It was found that the EBL flux seems to be lower than expected, thus allowing more distant VHE sources to be potentially detected. Furthermore the EBL seems to be dominated by galaxy population, mainly excluding stellar population~III contribution.



\begin{thebibliography}

\bibitem[Aharonian et al.(2004)]{2004APh....22..109A} Aharonian, F., et al. (H.E.S.S. Collab.)\ 2004, APh, 22, 109
\bibitem[Aharonian et al.(2005)]{2005A&A...442..895A} Aharonian, F., et al. (H.E.S.S. Collab.)\ 2005, \aap, 442, 895
\bibitem[Aharonian et al.(2006a)]{2006A&A...457..899A} Aharonian, F., et al. (H.E.S.S. Collab.)\ 2006a, \aap, 457, 899
\bibitem[Aharonian et al.(2006b)]{2006Natur.440.1018A} Aharonian, F., et al. (H.E.S.S. Collab.)\ 2006b, \nat, 440, 1018
\bibitem[Aharonian et al.(2006c)]{2006Sci...314.1424A} Aharonian, F., et al. (H.E.S.S. Collab.)\ 2006c, Science, 314, 1424
\bibitem[Aharonian et al.(2007)]{2007ApJ...664L..71A} Aharonian, F., et al. (H.E.S.S. Collab.)\ 2007, \apjl, 664, L71
\bibitem[Albert et al.(2007)]{2007ApJ...666L..17A} Albert, J., et al. (MAGIC Collab.)\ 2007, \apjl, 666, L17
\bibitem[Costamante \& Ghisellini(2002)]{2002A&A...384...56C} Costamante, L., \& Ghisellini, G.\ 2002, \aap, 384, 56
\bibitem[Fossati et al.(1998)]{1998MNRAS.299..433F} Fossati, G., Maraschi, L., Celotti, A., Comastri, A., \& Ghisellini, G.\ 1998, \mnras, 299, 433
\bibitem[Junor et al.(1999)]{1999Natur.401..891J} Junor, W., Biretta, J.~A., \& Livio, M.\ 1999, \nat, 401, 891 
\bibitem[Katarzy{\'n}ski et al.(2001)]{2001A&A...367..809K} Katarzy{\'n}ski, K., Sol, H., \& Kus, A.\ 2001, \aap, 367, 809 
\bibitem[Lenain et al.(2007)]{tmp_lenain} Lenain, J.-P., Boisson, C., Sol, H., \& Katarzy{\'n}ski, K.\ 2007, \aap, {\it submitted}
\bibitem[McKinney(2006)]{2006MNRAS.368.1561M} McKinney, J.~C.\ 2006, \mnras, 368, 1561 

\end{thebibliography}
\end{document}